# SMEs Confidentiality Issues and Adoption of Good Cybersecurity Practices

Alireza Shojaifar[1,2]

[1] FHNW, IIT, 5210 Windisch, Switzerland
`alireza.shojaifar@fhnw.ch`
[2]Utrecht University, Dept. of Information and Computing Sciences, Utrecht, Netherlands
`a.shojaifar@uu.nl`

**Abstract.** Small and medium-sized enterprises (SME) are considered more vulnerable to cyber-attacks. However, and based on SMEs characteristics, they do not adopt good cybersecurity practices. To address the SMEs' security adoption problem, we are designing a do-it-yourself (DIY) security assessment and capability improvement method, CYSEC. In the first validation of CYSEC, we conducted a multi-case study in four SMEs. We observed that confidentiality concerns could influence users' decision to provide CYSEC with relevant and accurate security information. The lack of precise information may impact our DIY assessment method to provide accurate recommendations. In this paper, we explore the importance of dynamic consent and its effect on SMEs trust perception and sharing information. We discuss the lack of trust perception may be addressed by applying dynamic consent. Finally, we describe the results of three interviews with SMEs and present how the new way of communication in CYSEC can help us to understand better SMEs' attitudes towards sharing information.

**Keywords:** Cybersecurity, Small and medium-sized enterprises, Privacy, confidentiality.

## 1   Introduction

Small and medium-sized enterprises (SMEs) as the backbone of the EU's economy [12] are the weakest targets for cyber-attacks [11]. SMEs have specific characteristics [13], and based on these characteristics, we are designing a Do-It-Yourself (DIY) assessment method, CYSEC, embedded in a tool. CYSEC wants to mitigate SMEs cybersecurity problems and support them to improve their security awareness and capabilities. CYSEC tries to establish a self-paced cybersecurity approach within a company. So, SMEs may select security focus areas at any time and based on their priorities.

Based on the first validation of CYSEC and observation of actual tool usage, we identified SMEs' confidentiality concerns as a barrier to answer self-reporting questionnaires and share their information. A reluctance to share security information may affect the quality of recommendations which are produced based on reported answers.

The consequence of non-practical recommendations may impact users' adoption of good cybersecurity practices.

The SMEs' behaviour may signify that they cannot trust the tool to keep their information confidential. So, we concentrated on the relationship between trust perception and confidentiality concerns. Therefore, and to impact the users' trust perception, a new privacy notice communication with the idea of dynamic consent has been added to CYSEC prototype to not only provide effective and meaningful notice for users but also give them the autonomy to change their preferences for sharing information at any time.

This study aimed to conduct qualitative research to (1) explore SMEs' cybersecurity responsible attitudes towards sharing security information with the CYSEC tool, SMEs community, and researchers and (2) to evaluate how the idea of dynamic consent model may affect SMEs decision to share their information and trust perception.

The remainder of the paper is structured as follows. Section 2 presents the CYSEC method, problem investigation, theoretical basis, and the research investigation. Section 3 describes the method of our study. Section 4 presents the analysis approach, obtained results, and the answer to the research question. Section 5 discusses the limitations of the study and the significance of the results. Section 6 summarises and concludes.

## 2 CYSEC Method and Problem Investigation

CYSEC is a DIY cybersecurity assessment method for SMEs. It automates elements of a counselling dialogue [2] between a security expert and employees in the SME to ward off cyber threats. CYSEC coaches SMEs to improve their cybersecurity awareness and capabilities through three Key features:

**Self-assessment Questionnaire**. Concerning a variety of security focus areas, CYSEC has self-reporting questionnaires. SMEs may select a focus area to answer based on their priorities.

**Training and Awareness Content**. Relevant training and awareness content are embedded in each questionnaire to not only indicate the threats or vulnerabilities but also demonstrate countermeasures against security threats.

**Recommendation.** A recommendation is a part of effective security communication between SME and cybersecurity expert. It provides training and awareness advice to prevent cybersecurity compromises and patch the vulnerabilities. The recommendations are produced based on users' answers to the questionnaires and may direct users to adopt good security behaviour.

For the first validation of the method, we conducted an explanatory multi-case study in four real SMEs, and the data collection was based on the observation strategy. During the studies, the researcher took notes about how the users applied CYSEC. Also, at the end of the actual usage of the tool, the users answered seven Likert scale questions and stated their attitudes.

The first validation revealed that confidentiality worries lead SMEs to avoid reporting their sensitive information or vulnerabilities. The researcher observed that the users were reluctant to answer specific questions such as "*Please list the systems that are open or offer shared accounts*" or "*Are the passwords sent encrypted between client and the server?*". One of the SME's security responsible stated: "*privacy issue [confidentiality] is necessary for our company any cybersecurity breach may severely threaten our business.*" Moreover, the SMEs' feedback and observations indicated that they are only worried about confidentiality and not the other security objectives: integrity and availability.

Advice and information are significant for SMEs to activate protective controls [8]. However, the users' reluctance to share their information with CYSEC and lack of accurate self-reporting information may affect the quality of the recommendations and advice. During the case studies and after replying to questionnaires, all the SMEs wanted to have a recommendation to indicate their next step. Therefore, personalized recommendations which indicate the SMEs' vulnerabilities and solutions can influence the users' adoption of good cybersecurity practices.

A reluctance to share security information is a kind of behaviour which can be explained based on users' relatedness to their data. In the context of security, users' emotional connection to their data demonstrates their perceptions of vulnerability and severity of security threats [5]. So, privacy concerns may influence people's willingness to disclose personal information [7]. Furthermore, privacy and trust may have a symbiotic relationship [7]. It means that "trust acts to moderate the impact of reduced privacy, whereas privacy also moderates the impact of reduced trust". So, privacy may be significant when there is a lack of trust in the requestor [7].

For the SMEs, we considered the relation between confidentiality and trust perception. When they do not trust the system to be able to keep the information confidential, they will not share their information. As a result, they would not receive strong recommendations.

### 2.1    Theoretical Background

CYSEC method considers features that support SMEs cybersecurity learning requirements through building intrinsic motivation and the possibility for selection between training options. Self-determination theory (SDT) concentrates on different types of human motivation and how humans' psychological needs for autonomy, competence, and relatedness should be satisfied for effective functioning and psychological health [4]. Autonomy refers to the choice of freedom and control over behaviour, competence refers to a feeling of achievement self-efficacy for an activity, and relatedness refers to a feeling of connection to others. These factors were considered in our case studies, and the author observed the relationship between these needs and the users' cybersecurity behaviour. Based on our observations, CYSEC needs to satisfy the users' autonomy, competence, and relatedness requirements to provide effective cybersecurity communication. These needs can be mainly operationalised through managing adaptedness of the self-assessment questionnaires based on users' answers, improving quality of content, and making connection among SMEs, security experts, and other SMEs.

## 2.2 Research Investigation

In the design of security systems, designers should consider the importance of humans in the loop. This means that designers of the system should recognise the potential causes of individuals' failure and support individuals in conducting their security-critical functions successfully [1]. Also, designers of security technologies need to understand how users make security decisions as well as the relevant security problems as a consequence of these decisions [9]. Based on SDT [4], relatedness is one of the significant users' psychological needs. To satisfy this need, mitigate users' confidentiality concerns, and motivate users to share their information with CYSEC, we made a change in the design of CYSEC prototype. The new design may improve our security communication approach to mitigate the confidentiality concerns and influence the users' trust perception.

Security communication may have different types, including warnings, notices, status indicators, training, and policies [1]. In case of information sharing communication, we considered notice type of message to help users to evaluate the effect of sharing information with CYSEC and make a decision. Notice alerts may be applied for the privacy policies. "Notices may be used by users to evaluate an entity and determine whether interacting with it would be hazardous or consistent with their security or privacy needs" [1].

In the design of our notice message and providing options for users to control their data sharing, the idea of Dynamic Consent in medical research has been applied. "Dynamic Consent uses information technology to facilitate a more explicit and accessible opportunity to opt out, whereby patients can tailor preferences about whom they share their data with and can change their preferences reliably at any time preventing any further data sharing" [10]. With this approach, the SMEs can see how their shared security data may be applied and how they can contribute to improving security issues in the ecosystem of cybersecurity.

Therefore, a notice box has been added to one of the questions in the CYSEC prototype to improve communication between SMEs and CYSEC. The main focus of the study was the content of the notice box and not the place of the notice. Fig. 1 shows a simplified page for a specific CYSEC question with a focus on the yellow notice box. The screen is split into two parts. On the left-hand side, the coach asks a question about the systems in the company with open or shared accounts, a box for the answer, and a yellow notice box. On the right-hand side, the training material for the current question is available.

The yellow box has three options with three goals and scopes to notify users of the data practices. SMEs are able to change their options and potential use of their data at any time. These options can show different levels of SMEs relatedness and data sharing. These levels are a) security experts, b) cybersecurity researchers, and c) the community of SMEs. In this study, we believe that a collaboration between all the groups may improve cybersecurity coping mechanism.

This research aimed to see 1) how the people responsible for the SMEs' security think about sharing security information and 2) if the new communication approach can affect SMEs decision, trust perception, and motivation for sharing their infor-mation. Therefore, four SME partners were invited to evaluate the new changes, and three of them were able to participate in three short semi-structured interviews.

**Fig. 1.** Screenshot of one of CYSEC questions with Dynamic Consent choices (yellow box)

## 3   Method

To achieve the aims of the study and consider the effect of new design on users' trust perception we conducted short semi-structured interviews [14]. The study wanted to answer the following research question:

*RQ1: How security communication through Dynamic Consent notice may influence SMEs confidentiality trust perception to share their information?*

To answer RQ1, SMEs partners has been invited for short interviews. Interviews were useful for allowing us to better understand SMEs mental model for data sharing. Each interview took maximum 15 minutes on Skype. In each interview at first the author presented concisely the relevant results of the first validation. All SMEs participated for the first validation and were aware of the problem. Then the author explained the concept of dynamic consent. Furthermore, the author presented the new notice box and its three options for sharing information. Table 1 presents our SMEs' demographics.

**Table 1:** SMEs Demographics

| ID | Industry type | Organisation size | Geographical distribution | Cybersecurity responsible | Implementation of cybersecurity controls | Structure |
|----|---------------|-------------------|---------------------------|---------------------------|------------------------------------------|-----------|
| 1 | Software development | Medium | Yes | Available | Some security controls | CEO, security team, employees |
| 2 | Software | Medium | Yes | Available | Some security | CEO, security |

| | | | | | controls | team, employees |
|---|---|---|---|---|---|---|
| | development | | | | | |
| 3 | Software development | Small | Yes | Available | Some security controls | CEO, security team, employees |

## 4 Analysis and Results

The interviews' topic was developed based on the sharing information problem which identified in the first validation and the idea of dynamic consent model. The users explained their attitudes about dynamic consent and its effect on their trust perception. However, during the interviews, the users indicated other issues rather than trust perception, and the author considered them during the code development phase. So, two significant themes were determined to characterize the users' attitudes towards the CYSEC new design and dynamic consent:

**The role of trust perception and the level of sharing information agreement (RQ1).** The users indicated different degree of trust after considering the dynamic consent choices. *SME #1* select the first option to share the data only for receiving recommendations and stated that "*this box improved my trust to the CYSEC since it can show that the tool considers the confidentiality and there is a process to protect the data*". With respect to the other options the *SME #1* explained that "*I can select option three if there is an agreement for sharing information with community*".

This selection approach was almost the same with the *SME #3*. *SME #3* explained that "*this design improved my trust. Now I can decide how the tool can apply my data*". The user of company #3 also stated that "*as a SME I will select option 1 because it is safer, however, as a use case partner I will select option 3 to help the community*".

In contrast, SME #2 did not select any options of dynamic consent and explained that "*I think answering critical questions has no value to our company. For critical questions I prefer to use training content [right-hand side of the CYSEC coach] to understand for instance, why sharing accounts are not acceptable and how to stablish a solution*".

Although two of the SMEs indicated that the new design of the CYSEC could influence their trust perception to share the information with the tool, none of the participants selected the second option to collaborate with security researcher. Also, they emphasised the importance of an agreement for information sharing with community.

**Dynamic consent and operational issues.** SME #1 and SME #3 indicated some issues about the dynamic consent box. Both explained that it is easy to use, however, it is better to have the box in the beginning of each coach and it is annoying that we see the box several times for some questions. None of the added more options to the list while SME #1 wanted a change in the order of second and third options. In addition, SME #1 stated that "*I assume that the tool should destroy the information after generating recommendations*".

## 5 Discussion

### 5.1 Study Limitations

Before conducting the study, the author presented the design prototype and dynamic consent options to a cybersecurity expert to validate the options. Also, each interview was conducted without any break. However, our study has certain limitations, some of which the author hopes to improve upon in future work. First, the number of interviews in this research was only three. Also, we did not use different data sources to consider the triangulation effect for validity. In addition, the number of questions for the interview was not enough to study various aspects of the users' point of views.

### 5.2 Cybersecurity Information Exchange

Our study indicated that having an agreement for sharing security information may encourage SMEs to collaborate in cybersecurity improvement. Therefore, planning an agreement with SMEs for sharing their security information may support the cybersecurity community and coping mechanism. The study showed that SMEs needs to understand more about the security collaboration with community and researchers to not only comply with security recommendations but also adhere to good cybersecurity practices. Still stimulating organisations to participate in security information sharing is a challenge. Rewarding, participation-fee, and economic incentives may be applied for achieving the company's participation [15] [16].

## 6 Conclusions and Future Works

The paper has evaluated the effect of applying dynamic consent model on SMEs' trust perception to share their cybersecurity information with CYSEC method. CYSEC is a do-it-yourself (DIY) security assessment and improvement method for small and medium-sized enterprises (SMEs). In the first validation of CYSEC, we understood that confidentiality worries lead SMEs to avoid reporting their sensitive information. This behaviour may affect the quality of CYSEC recommendations, and non-practical recommendations may impact users' adoption of good cybersecurity practices.

To evaluate our new design, the dynamic consent approach has been implemented with three options to share the information. The analysis of the results demonstrates that the dynamic consent box has a positive effect on some of the SMEs trust' perception to share their information with CYSEC. However, they were reluctant to share their information with community and security researcher. Sharing the information with CYSEC has a value for SMEs to receive recommendations. However, it seems that the other options have no incentives for the SMEs.

As future work, we have planned to invite eight new SMEs to evaluate the effect of the new design of CYSEC.

## Acknowledgments

This work was made possible with funding from the European Union's Horizon 2020 research and innovation programme under grant agreement No 740787 (SMESEC) and the Swiss State Secretariat for Education, Research and Innovation (SERI) under contract number 17.00067. The opinions expressed and arguments employed herein do not necessarily reflect the official views of these funding bodies.